# Large Language Models as Instruments of Power: New Regimes of Autonomous Manipulation and Control

Yaqub Chaudhary[1]    Jonnie Penn[2]

3 May 2024

Abstract: Large language models (LLMs) can reproduce a wide variety of rhetorical styles and generate text that expresses a broad spectrum of sentiments. This capacity, now available at low cost, makes them powerful tools for manipulation and control. In this paper, we consider a set of underestimated societal harms made possible by the rapid and largely unregulated adoption of LLMs. Rather than consider LLMs as isolated digital artefacts used to displace this or that area of work, we focus on the large-scale computational infrastructure upon which they are instrumentalised across domains. We begin with discussion on how LLMs may be used to both pollute and uniformize information environments and how these modalities may be leveraged as mechanisms of control. We then draw attention to several areas of emerging research, each of which compounds the capabilities of LLMs as instruments of power. These include (i) persuasion through the real-time design of choice architectures in conversational interfaces (e.g., via "AI personas"), (ii) the use of LLM-agents as computational models of human agents (e.g., "silicon subjects"), (iii) the use of LLM-agents as computational models of human agent populations (e.g., "silicon societies") and finally, (iv) the combination of LLMs with reinforcement learning to produce controllable and steerable strategic dialogue models. We draw these strands together to discuss how these areas may be combined to build LLM-based systems that serve as powerful instruments of individual, social and political control via the simulation and disingenuous "prediction" of human behaviour, intent, and action.

Introduction

In January 2023, conveners of the International Conference on Machine Learning (ICML) banned the submission of text produced entirely via large language models (LLMs). Their ban did not prevent "editing or polishing author-written text" with such tools (ICML Program Chairs, 2023). This exclusion suggests that even in a setting in which experts deem LLMs to be a risk, little is made of their use in crafting composition and rhetoric. We challenge that assumption herein by surveying LLM capabilities that, to our knowledge, have not yet been examined systematically elsewhere.

[1] Visiting Scholar, Leverhulme Centre for the Future of Intelligence, University of Cambridge
[2] Research Fellow, Leverhulme Centre for the Future of Intelligence, University of Cambridge

In what follows, we outline several reasons why the use of LLMs to alter writing styles is controversial, particularly in areas of sustained human-machine interaction. Sustained interactions tend to be understood through the lens of user prompts. We add to this that LLMs prompt *back*. In other words, LLMs equip powerful actors with new capacities for the projection of power over users. We caution that the scope of this power expands in proportion to the size of the context that an LLM operates within (or upon). In the case of Microsoft and other tech multinationals, that context is global.

This paper considers the incentive structures and political possibilities latent in that global scale. That the development of LLMs is deeply ensconced with the prerogatives of the advertising industry (Whittaker, 2021) raises significant questions over their proposed uses in healthcare, education and in the workplace, given the problematic associations between advertising, marketing, public relations and propaganda (Marlin, 2002: 13; Zollmann, 2017). Existing scholarship shows that the dissemination of deceptive content and disinformation is incentivised by the market structures within which digital platforms operate (Diaz Ruiz, 2023); a finding that invites critique over how LLMs will be used to influence, persuade, and control wide audiences.

In the same month as the ICML ban, LLM-developer OpenAI was valued at an estimated USD 19-29 billion, ranking it one of the most valuable start-ups in America (Kruppa, 2023). The basis of this steep valuation may be understood by considering the plans of its industrial benefactor Microsoft, which in February 2023 announced that it intends to create "new value for advertisers" via the integration of LLMs into its Bing search engine. Microsoft seeks to re-architect computing infrastructures to make LLMs, foundation models and generative AI essential components of computing platforms and services. At the first OpenAI developer conference, Satya Nadella remarked that the rise of LLMs is prompting them to re-think "the system, all the way from thinking from power to the DC to the rack, to the accelerators, to the network" (OpenAI DevDay, Opening Keynote, 2023).

One reason the major technology companies are incentivized to interweave LLMs throughout the web is that doing so positions them as platform intermediaries between users and the web. Jensen Huang, CEO of NVIDIA, casts AI as "a new computing platform" that will re-configure online environments into generative feeds of information wherein users are co-opted into a "dialogue" with web resources and information services (NVIDIA Keynote at SIGGRAPH 2023, 2023). According to Huang, "The canonical use case of the future is a large language model on the front end of just about everything. Every single application, every single database, whenever you interact with a computer, you will likely be first engaging a large language model" (NVIDIA 2023).

In assessing LLM-related societal harms alongside economic and infrastructural realities, we break with literature that foregrounds their use to disrupt or pollute information channels and alter social norms. Instead, we focus on how such disruption and pollution fits into broader efforts to centralise power by expanding the surface area available for social control. A first wave of critique noted the displacement of journalists and writers (Naughton, 2022), the "destabilisation" of white-collar work (Lowrey, 2023) and related changes in education (Marche, 2022). *The New Yorker* and *Wired* magazine suggest, alongside academics, that language models are "poisoning web search" (Chiang, 2023; Knight, 2023) with "semantic garbage" (Floridi and Chiriatti, 2020). These framings foreground how algorithmic tools alter the character of the setting(s) in which they operate in. We, in contrast, are interested in how LLMs can be used not just to operate *in* a given context, but also to operate *on* it using automated strategies to deliberately manipulate users to elicit a particular political end.

In existing literature that gets closest to this topic, Weidinger et al. (2022) distinguish between LLM's "observed" and "anticipated" risks. Although we do not invoke this terminology here, our analysis entails both categories, meaning *observed risks* (wherein existing empirical research has identified LLMs' manipulative effects) and *anticipated risks* (which theorises new manipulative mechanisms latent in emerging systems). Since Weidinger et al.'s useful taxonomy was published prior to the release of ChatGPT, GPT-4 and the AI agent, Cicero, their binary is unfortunately already outdated. So-called anticipated risks are already "in the wild," or are fully actionable, rather than being observed for their potential. We thus build on their analysis.

In summary, this article attempts to understand key aspects of the incentives driving ambitions to make LLMs fundamental mediators of human-computer interaction. In doing so, we reject the disjunction suggested by the designation of large language models as so-called "foundation models" and re-situate the field of research and applications they have spawned as in-keeping with prior AI research premised in control, such as deep reinforcement learning in adversarial games undertaken in the 2010s. In this connection, we focus primarily on Cicero, the AI agent from Meta Platforms, Inc. that exhibits human-level play in the game Diplomacy using LLMs and reinforcement learning to craft strategic and adversarial rhetoric.

## Large Language Models as Instruments of Power: Techniques for the Projection of Power and Control

LLMs can reproduce a wide variety of rhetorical styles and generate text that expresses a broad spectrum of sentiments. This capacity, now available at low cost, makes them powerful tools for rhetoric and persuasion. In this section, we identify four ways in which humans will interact with the persuasive capabilities of LLMs:

1. Via changes to digital environments as a whole (ex., by changing internet use)
2. Via a direct site of interaction between a single human and an LLM system (ex., use of ChatGPT)
3. Via a proxy of the user trialled in predictive models and simulations to model user intent contextually (ex., use of Cicero)
4. Via multiple proxies operating within and/or around a limited social sphere (ex., within an online community containing interactions between multiple human and machine agents, or a simulation of an office space, voting district, transit system, etc.).

These four areas are neither mutually exclusive nor wholly distinct—indeed, it is their coordination and combination that concerns us. To illustrate that these scenarios are not hypothetical, we end this section by considering how Cicero represents an important proof-of-concept for the strategic control and steering of human subjects in digital environments. Before considering such overlaps, however, we will introduce each of the four areas in turn.

### (i) LLMs and Digital Information Environment as a Whole: Pollution and Uniformization of Content and Experience

From the 1990s to the 2020s, the internet was populated largely by media created by human contributors. Generative AI and LLMs alter this norm in ways that we can only begin to characterise here. As a fundamentally unstable form of software capable of behaviours that cannot yet be formally verified[3] (Sommart, 2023: 6) or "explained" using established approaches from mathematics, philosophy, or computer science, LLMs deepen our dependencies on "incomplete" (Bommasani et al., 2022: 1, 7) computational instrumentation with unknown failure modes and outputs. Poor training data exacerbates this problem, as does the possibility that "defects of the foundation model are inherited by all the adapted models downstream" (Bommasani et al., 2022: 1). We focus here on the "pollution" and countervailing "uniformization" of semantic content in informational environments exposed to such systems.

(a) Pollution

LLMs can be used to produce large quantities of intelligible text at low cost. Scholars equate this capacity to "pollution" because it compromises the veracity of critical information channels like the internet. At worst, this capacity could actualise the unending Library of Babel described in the short story of Jorge Luis Borges, which contains, "the translation of every book in all languages, the interpolations of every book in all books" (Jorge Luis Borges, n.d.). The sheer size of this library makes it impossible to read or learn anything meaningful because the most precious books are forever out of reach. While feelings of "information overload" predate the digital era (Blair, 2010), the low-cost production of generative media threatens to add industrial scale to the obfuscation of meaningful information online. Deepening this problem is the prospect that LLMs be positioned as the tools needed to *remediate* retrieval (Zittrain, 2022), as we discuss in the next section.

This "pollution" phenomenon is not constrained to the internet. The applied machine learning researcher John Nay demonstrates a proof-of-concept for the use LLMs in corporate lobbying. Nay's model sifts through US Congressional bills for relevance to specific companies. Where a bill is found to be relevant, the model (GPT series) drafts a persuasive letter to its congressional sponsor to make changes to the proposed legislation in favour of the company. According to Nay, AI lobbying may lead to a slow drift in information on values held by citizens, such that policies cease to reflect their preferences in favour of corporations or other powers (Nay, 2023).

The gradual accretion of semantic garbage may be regarded as a form of slow violence (Nixon, 2011) that destroys beneficial information by poisoning the information environment as a whole. Bommasani et al. argue that LLMs can be used to collapse the variety and diversity of information online through the imposition of an "algorithmic monoculture" that acts as an "epistemically and culturally homogenizing force" to spread "one implicit perspective, often a socially dominant one, across multiple domains of application" (Bommasani et al., 2022: 1). Given uncertainties about their failure modes, LLM content could propagate information riddled with unintentional errors. Alternatively, LLMs could be used as a scaffold for intentional falsehoods hidden amongst true information. In sum, LLMs "polluting" influence on informational spheres remains a critical open question.

(b) Uniformity

---

[3] "strong, ideally provable, assurances of correctness with respect to mathematically-specified requirements," (Seshia et al., 2020).

If "pollution" captures how LLMs can be used to diffuse information environments, "uniformity" captures how LLMs can be used to assign *order* to them, such as the order desired by an institutional power. History makes prior efforts in this direction legible in ways that present-day critiques overlook, since they occur over small and large timescales. The historian Loraine Daston attributes the conformity of public experience over past centuries to European empires' enforcement of standardised rules, including rules for language. She writes:

> Trends, such as the increasing rationalization of work in industrialized societies, as well as ideals of natural law imported from theology into natural philosophy and from there into jurisprudence and ethics… promoted rules of ambitious globality and exactitude. Proliferating especially (but not exclusively) in urban settings in the modern era, these rules appeal ever more to universal principles (whether of the market or the rights of man) and ever less to local context and background knowledge. Not coincidentally, the rise of such ambitious rules begins with the expansion of trade and empire to global dimensions in the sixteenth century, which created both the need for and the means to enforce rules that transcended any one locality (Daston, 2022: 19).

For our summary in this sub-section of how the rhetorical capabilities of LLMs assert uniformity on to social experience, we draw on research by the linguist Uwe Poerksen, who argues that the politics of language manifest over long time horizons. Poerksen attributes global uniformity of experience today to the politics of language during the French revolution, in which, "Making language uniform was part of an attempt to achieve uniformity in general" (Poerksen, 1995: 33). Poerksen describes how "information," "development," "progress," and "communication" have been shorn of their precise meaning in scientific lexicon and repurposed with a high degree of *plasticity* in everyday language. Many of these words maintain the authority of their scientific origins even while their resemblance to science is superficial. They "transmit the reputation of science into the everyday," Poerksen writes, and in ways "that disable the vernacular" (1995, 75, 5).

Importantly, such words are plastic in another sense—that of the plastic Lego blocks used to build structures. LLMs provide the componentry for a wholesale re-ordering of the built environments that structure our perception of reality. "Amorphous plastic words are the elemental building blocks of the industrial state," Poerksen argues, and are used by politicians, bureaucrats, consultants, industrialists, academics and others as "ciphers" to "clear the way for operations on a grand scale" (1995: 5). This need not be clandestine to be effective in drawing attention away from certain aspects of social change. Architectural design historian and theorist Keller Easterling argues that the alteration of built environments is often framed in other terms "to disguise or distract from what the organization is actually doing" (Easterling 2016).

While the uniformization of language in the history of European colonialism has been a key to creating the conditions for effective social control, LLMs appear to invite new levels of social conformity. Poerksen positions the digital computer as "the consequence" and complimentary expression of centuries-long campaigns to make language uniform in bureaucratic, corporate and political texts (Poerksen, 1995: 91). Equally we might say that LLMs are the *next* consequence and logical expression of such traditions, insofar as they render humans as "victims of a totalitarian monism of natural sciences and their technological elaborations" (Poerksen, 1995: 98). LLMs extend the capabilities for refashioning language in the image of plastic words; they reduce text to tokens in ways that can be recombined and recombined *ad nauseum*. LLMs inherit a legacy of making language an instrument for social control through the uniformization of language and thus underlie a new form of power for restructuring reality through text, and of course, code, and any other aspect of reality that may be tokenised.

(ii) Direct Human-LLM Interaction with Individuals: Persuasion through Choice Architectures and Conversational Interfaces

The field of persuasive technologies aims to design tools that change attitudes and influence behaviours of users, not only via persuasion through text, but also via interfaces that entice users toward designers' preferred choices and outcomes within what is called a "choice architecture" (Yeung, 2017: 120). LLMs provide new means to order and structure choice architectures in online environments, both in single sessions and between sessions. They do so by automating the personalisation and persuasiveness of interfaces, as well as by persuading more literally through the use of conversational agents. Mills and Sætra describe this as "autonomous choice architects" (2022).

The advent of "AI personas" offers one early example of conversational interfaces becoming a vehicle for persuasion. The concept is a key ambition for Meta Platforms, Inc., which in 2023 released a series of products that integrate generative persona interactions into WhatsApp and Instagram (Meta, 2023). In this product category, the manipulative capabilities of an LLM interface can be augmented both by the generation of persuasive text, and/or by masquerading in the persona of someone who may be considered a trustworthy source by the user. Personalised chatbots under these guises may be used to persuade users on behalf of commercial or political campaigns in a one-to-one setting (Goldstein et al., 2023: 25).

A new body of literature explores the theoretical foundations of persona simulation using LLMs (Jiang et al., 2023; Serapio-García et al., 2023). Serapio-García et al. attempt to provide an empirical framework to quantify LLM personality through the same psychometric methods that are applied to humans. They argue that such quantification paves the way for personas to be "verifiably shaped along desired dimensions," wherein "shaping" involves mechanisms to modulate (e.g., increase or decrease) "levels of specific LLM personality traits" (2023: 14). They position this research as an exercise in translation, one that re-directs "established measurement theory from quantitative social science and psychological assessment" to the study of LLMs and toward, ultimately, a "science of LLMs".

Serapio-García et al. draw attention to potential misuses of LLMs personas. They note the possibility of personality matching (e.g., matching content to a user's personality), which has been found to be highly effective at enhancing levels of persuasiveness and influence (Matz et al., 2017; Tapus et al., 2008). Interestingly, the authors find that small optimised LLMs are also able to reproduce complex personality profiles, an advent that lowers the barrier, in terms of computational resources, for the mass, low-cost deployment of LLM-based agents with synthetic personalities. We add to this that the manipulative and persuasive capabilities of LLMs do not need to be manifested through conversational agents to be effective. The integration of LLMs into real-time bidding (RTB) advertising networks, for instance, provides new opportunities for influencing users' thoughts, opinions and preferences via conversational interfaces, such as by inserting advertising content based on the mention of certain key words, or by positioning an LLM as the spokesperson for a certain brand.

In this sub-section, we have introduced several ways in which LLMs may be used to generate persuasive content at the level of the text itself, via autonomous choice architectures, and by exploiting the tendency of humans to anthropomorphise computational agents, such as through AI personas. Next, we consider mechanisms of influence in the co-production of text when LLMs are used for predictive composition.

### (iii) Targeted Human-LLM Interaction using Mirroring and Steering: Interposing an LLM-agent as a proxy for the user in predictive modelling and simulation

In this section we explore a heightened set of examples of LLM-based persuasion. Specifically, we explore how existing predictive modelling and simulation techniques can be used to be more deliberately persuasive than the techniques covered in the previous section. An LLM-system might, for instance, be purposed as a predictive model that is fine-tuned on the user's own writing such that its outputs are aligned to the writing style of the user, tempting out a sense of familiarity. In this way, LLM's use in suggesting composition options may subtly undermine a user's agency by appearing to provide suggestions that the user *feels like* they would articulate themselves. We refer to this broad category of influence as "mirroring and steering."

**(a) Mirroring and steering dialogue through infiltration**

In human relationships, the use of obsequious behaviour to gain advantage is called sycophancy. The advent of sycophancy in advanced conversational agents raises questions about links between language and thought. Various theorists have advanced that "language production and comprehension are tightly interwoven" (Hancock et al., 2020; Pickering and Garrod, 2013). The questions we consider here pertain to whether predictive compositions influence linguistic expressions exclusively, or whether there is a deeper level of influence that shapes what a user actually thinks in the process of writing. Each instance merits closer study. Here, we survey various modes of interaction whereby the user may voluntarily discard their own manner of self-expression of their thoughts for that of the LLM.

Jakesch et al. (Jakesch et al., 2023) have demonstrated that an LLM may be repurposed for predictive composition. They designed an experimental system that uses a pre-configured LLM to generate tokens after a user pauses writing. We anticipate that a more advanced system could employ an active predictive model that prompts suggestions rather than waiting for the user to pause. Such a system could update itself dynamically to be a few words ahead of the human writer or speaker. As a simple analogy, imagine auto-complete text designed to persuade its user to write in a certain way. What we wish to highlight here is that if what someone will say next in a conversation can be anticipated with a high degree of confidence, it becomes possible to manipulate the overall flow of that conversation. Research that pre-dates the major developments in LLMs shows that the more a user follows predictive suggestions, the more predictable the content of what they are writing becomes (Arnold et al., 2020: 128). A related scenario is for the user to simply accept a complete passage of text generated by an LLM instead of their own writing.

**(b) Steering via "snap to grid" thinking**

"Predictive" completion hinges on the willingness of individuals to surrender their own thoughts to that of the predictive system. Crucially, the surrender of thought opens the way to the steering of thought, which may be enabled by the integration of systems for controlling LLM outputs such as the AI agent, Cicero, which as we discuss later in this article, can be adapted for strategic dialogue with and the strategic prompting of a user. Studies in human-computer interaction have shown how agency is undermined by psychological techniques such as computer-assisted movement on screen. For example, in one experiment involving a point-and click task with varying levels of computer assistance, it has been shown that, up to a certain level of computer assistance, users maintained a sense of agency over their actions (Coyle et al., 2012: 2030–2034). In principle, this influence can be modulated in real-time to express certain sentiments or even to align the content of prose to specific political or ideological axes by making suggested text visible in line with manual composition. The modulated production of text as it is being composed may be viewed as analogous to the "snap to grid" mechanism that appears in graphic design applications. However, instead of snapping to a visual grid, ideas, attitudes, and beliefs expressed through text are snapped along various axes in high dimensional spaces captured by neural networks.

Here, we note that there is a two-way modulation that arises from the mass production of synthetically generated text (which ultimately rests on prior human composition) alongside fresh text of purely human compositions and text that is freshly co-produced by human-LLM interaction. This mixture creates the conditions of possibility for accentuating recursive shaping of opinions and attitudes in online social settings. First, by adding to social pressure experienced by vast numbers of people for performative exposition of their private lives in public forums, and second, by multiplying re-conditioned versions of this content which gradually aligns human contributions to the standards channelled via LLMs. While mirroring the behaviour of a pack is not a new or unnatural phenomenon, LLMs allow for subtle modes of intervention in the re-setting of norms.

**(c) Archetype mining: sampling and simulating LLMs as "silicon subjects"**

This sub-section surveys concurrent developments in the computational social sciences (Argyle et al., 2023), political science (Ornstein et al., 2023) and economics (Nay, 2023), as well as psychology (Demszky et al., 2023), which provide methods for experimenting with LLMs as a new kind of scientific object within these disciplines. These new methods enable scientists to conduct social science studies on populations of synthetic individuals generated through prompts. Without LLMs, this type of study would require tremendous resources and research time on *actual* living persons and collectives. Taken together, these methods contribute significantly to expanding the possibilities, effectiveness, and cost-effectiveness of LLM-based manipulation. In addition, these developments in the scientific use of LLMs converge with aforementioned research on the generation, evaluation and modulation of LLM-based personalities in ways that will be discussed later in this article. Broadly speaking, this prospect is a logical follow-on of two-centuries of consolidation in the institutional use of data doubles (Bouk, 2017; Haggerty and Ericson, 2000).

Argyle et al. have introduced the concept of "silicon subjects" to the social sciences based on an insight into the nature of LLMs that runs contrary to the prevailing discourse on "algorithmic bias" that positions bias as a deficiency to be mitigated. In contrast, they argue that biases are "demographically correlated" such that LLMs can used to reproduce "response distributions" from human subgroups, and are thus capable of standing in as "surrogates for human respondents in a variety of social science tasks" (Argyle et al., 2023: 337). Argyle et al. describe this ability to treat algorithmic bias as a proxy for demographics as "algorithmic fidelity". Algorithmic fidelity naturalises the idea that there is value in conducting a census on "silicon subjects," since data on users' speech characteristics, for example, can be used to understand, map, and relate different clusters of a given population.

Practitioners in the social sciences and in industry have begun to adopt this approach, often uncritically. Horton explores using LLMs as "simulated economic agents' and argues that such agents represent "implicit computational models of humans" (Horton, 2023). In a pre-print paper, he argues that GPT-3 level models (and above) have introduced a categorically new type of experimental tool for studying computational models of human subjects. Elsewhere, Brand et al. propose to use LLMs for market research, such as in marketing and pricing strategies, to elicit willingness-to-pay measurements from GPT-3 responses, which can serve as a proxy for consumer preferences (Brand et al., 2023). Outside of academia, industrial actors reference this new area of research when commercialising analogous claims. The Portuguese start-up Synthetic Users professes it can "accelerate" user and customer research activities via LLMs tested against samples of marketing text and they appeal to Argyle at al. as providing the empirical foundations for their product placement services (Synthetic Users, n.d.).

The scientific basis of these prospects is unclear. Park et al. identify issues in the replicability of LLM-as-proxy studies, such as the "correct answer" effect. This effect refers to a tendency of the GPT 3.5 model to answer subjective and nuanced survey questions as though there is only one correct answer (PS Park et al., 2023). Eric Chu et al., who use Google's BERT model to predict political opinion from different patterns of media consumption, caution that whilst language models can predict public opinion and human survey responses, their result do not imply that human participants or surveys can be substituted by AI models (Chu et al., 2023: 9). Instead, they, alongside others such as Rosenbusch et al., argue that LLMs may provide a reference for how useful certain types of real-world studies may be, or serve as, a creative tool for planning research and generating hypotheses (Rosenbusch et al., 2023: 17).

As mentioned, whether or not such functions are scientific is a different matter from whether they will be used. The cost of the simulations by Argyle et al. ($29 USD on GPT-3) are markedly lower than traditional methods used to identify operative words and relationships in a given social setting (ex. via focus groups) and Argyle et al. note that such tools "could be used to target human groups for misinformation, manipulation, fraud, and so forth" (Argyle et al., 2023: 349). Adding to this problem is the fact that an estimated 33-46% of crowd workers have used LLMs for abstract summarisation tasks (Ornstein et al., 2023; Veselovsky et al., 2023). This creates a vicious cycle that diffuses the compounding role of LLMs on media environments.

Overall, we argue, the legitimacy of "silicon subjects" as a proxy for human subjects deserves close attention to avoid negative outcomes. Even if the new LLM methods are not as efficacious as existing techniques like focus groups, and despite their being prone to high error rates in many situations, concerns over their use stem from the fact that they can be applied at low-cost and at scale. In 2023, Ornstein et al. showed that replicating a 2017 study that coded 935 political ads (Carlson and Montgomery, 2017) took less than two minutes using an LLM and cost only $18.46 USD compared to $565.20 in the original study, yet produced results that were more strongly correlated with expert ratings (Ornstein et al., 2023). The most obvious concern over these prospects relates to the future of automated propaganda. A less obvious concern is that the body of purportedly scientific research that becomes established through such methods might come to influence policy and law that reflects the responses of synthetic populations in favour of real people over whom such policies are to be applied. Further concerns emerge in consideration of the ease of scaling the number of LLM instances, which paves the way to new multi-agent social simulations calibrated with user data, which we discuss in the following section.

(iv) Multi-Archetype Mining: Trialling "Silicon Societies" using Multiple Silicon Subjects

As discussed in section (iii), there is increasing interest in the possibility that appropriately conditioned LLMs can be studied as proxies for human participants in social, political, and economic studies (e.g., "silicon samples"). A logical extension of this developing work on LLMs as computational human models is to progress from the simulation of individuals to the simulation of social interactions via the use of synthetic populations of individuals, which we'll call "silicon societies." The extension of LLM technology in this direction would introduce new capabilities for manipulating and controlling social discourse across various scales.

The prospect of using LLMs to generate agents in social simulations was first demonstrated soon after the public release of ChatGPT, illustrating how quickly modular techniques in AI can be combined to achieve new functionalities (JS Park et al., 2023). Park et al. devised an architecture consisting of an "interactive sandbox environment" populated by computational agents, which are individually generated through prompts to ChatGPT in order to produce a "small society" of twenty-five agents. The LLM stores "a complete record of the agent's experiences", synthesizes them over time into "higher-level reflections", and then retrieves them to "dynamically plan behavior". The authors claimed that their simulation yielded emergent phenomena such as information diffusion across agents, memory of relationships between agents, and coordination of social interactions amongst agents. Further, they claimed such agents are destined for action beyond a sandbox environment, and this research "opens up the possibility of creating even more powerful simulations of human behavior to test and prototype social systems and theories" (2023: 17). In another 2023 study of social simulations using 31,764 LLM-based agents, the authors claim to have found "early evidence of self-organized homophily in the sampled artificial society" (He et al., 2023: 11). This study used Chirper.ai, which is a platform resembling Twitter that is exclusively for LLM chatbots. Such research illustrates new possibilities for comparisons between the collective social behaviour of LLMs and the collective behaviour of humans, as well as the possibility of using LLMs for more advanced methods from the social sciences in the context of collective social behaviour.

In her discussion on the role of models in economics, Mary Morgan discusses how models have increasingly come to serve as instruments for acting in the world (Morgan, 2012: 400). According to Marx Wartofsky, models represent a "mode of action" and that they are "embodiments of purpose and, at the same time, instruments for carrying out such purposes" (Wartofsky, 1979: 142). We argue in this context that the use of LLMs for simulating silicon subjects and computational modelling of "silicon societies" foreshadows their use as instruments for intervening on and controlling human behaviour and psychological dispositions, as we will now discuss via a focus on the Cicero model.

## "Cicero" as a Proof of Concept for Strategic Dialogue: Control and Steer Dialogue by Combining LLMs with RL

In this section, we argue that Meta's AI agent "Cicero," first announced in November 2022, represents an important proof-of-concept for the sorts of strategically personalised messaging outlined above. Cicero combines LLMs, real-time services, and reinforcement learning (RL) to steer human subjects in digital environments. The historical figure Marcus Tullius Cicero (d. 43 BC), after whom the system is named, wrote extensively on the power of rhetoric and persuasion, particularly in the domain of statecraft. In *De Oratore*, he described the ideal orator as the ideal moral guide of a state and cautioned that an orator ungrounded in moral principles could use their persuasive powers to manipulate a community in support of their own personal ambitions. We speculate that the name "Cicero" suggests an ambition on the part of Meta to either provide strategic dialogue services directly to business users running commercial, political, or other forms of influence campaigns or to facilitate the deployment and use of such agents across its services[4].

Developers at Meta claim that Cicero (the model) is the first AI agent to have "mastered" the skill of using language to "negotiate, persuade, and work with people to achieve strategic goals" (Meta AI, 2022). They make this claim via demonstrations of its capacity for human-level play in the strategy game Diplomacy (Bakhtin et al., 2022; Meta AI, 2022). Diplomacy is a board game in which players undertake several rounds of negotiations to advance their position on the board by forming alliances with other players or by betraying them conversationally. Unlike chess or Go, wherein success is achieved by efficiently computing a game state in relation the value of possible moves, success in Diplomacy necessitates modelling the *hidden intents* of other players. Together, these features make Diplomacy an exemplary microworld that can be readily mobilised in analogous fashion to a wide range of domains and contexts where the aim is to persuade a counterparty to undertake a desired course of action, be it purchase a product or vote for a political candidate.

In principle, Cicero supersedes earlier techniques of "digital mass persuasion" (Matz et al., 2017) which relied on single parameters such as Facebook Likes to predict personal attributes (Kosinski et al., 2013). Whilst there has been scepticism over the science and success of applying these and other methods of psychological targeting, especially in the context of the Cambridge Analytica scandal (Gibney, 2018; Sharp et al., 2018), Matz et al. caution that automated content generation with LLMs could make personalised persuasion scalable, more effective, and more efficient (2023). In other words, the history of persuasion did not end with Cambridge Analytica (hereafter, CA). One of the most significant challenges this raises, as previously noted by Benkler et al., is that "behaviourally informed, microtargeted dark ads are likely the most important novel threat to democratic practice" (Benkler et al., 2018: 223). Given the possibilities fulfilled by LLMs on persuasive capabilities not achieved by CA, we will now characterise three ways in which the use of models like Cicero could amplify persuasion online.

**(i) The elicitation of private data**

The first area we consider involves optimising methods for the elicitation of private data. On this topic, Matz et al. argue, "The effectiveness of large-scale psychological persuasion in the digital environment heavily depends on the accuracy of predicting psychological profiles" (2017, 4). Accordingly, an actor wishing to engage in digital mass persuasion would be best availed by being able to access a "full history of digital footprints" in order to "continuously calibrate and update" (2017, 4) their algorithms over time. As auto-regressive models, LLMs facilitate such efforts. Staab et al. show, for instance, that LLMs can infer personal information through seemingly benign conversational exchanges, and can even "steer conversations" in such a way as to provoke responses from which to infer private information (Staab et al., 2023).

---

[4] We note that the corresponding author for this work, Noam Brown, joined OpenAI in mid-2023, which we suggest signals an ambition on the part of OpenAI to integrate strategic dialogue capabilities in future LLM services.

The latent potential of existing surveillance regimes adds gravity to this ability. Prevailing social media infrastructure and other large-scale digital repositories (e.g., e-commerce, e-governance, e-medicine, etc.) combine both public and private data collection in ways that have not yet been "tapped" with the level of fidelity made possible—or, rather, affordable—via the introduction of LLMs. Computer historian Luke Stark traces prior investments in, and testing of, techniques to make human subjectivity calculable via the production of "scalable subjects" (Stark 2018). He cites, for instance, Facebook's "emotional contagion" study in 2014 and efforts by Facebook's Australian division to make algorithmic psychometrics legible for advertisers so that they could target "teenagers based on real-time extrapolation of their mood," (Stark, 2018: 206; Tiku, 2017). To understand how access to private data enables attempts at persuasion, we will now turn to a second related capability made feasible using LLMs—and Cicero in particular.

**(ii) The strategic personalisation of generated content**

Central to the novelty of Cicero and what differentiates it from other language models is the concept of a "controllable dialogue model" (Bakhtin et al., 2022). The base model of Cicero, called R2C2, is a relatively small transformer model with only 2.7 billion parameters, compared to hundreds of billions in the state-of-the-art LLMs of the mid-2020s. The relation between Cicero and much larger models such as ChatGPT is that the latter delivers probabilistic responses to each prompt (e.g., attuning to the so-called "temperature" of each output). Cicero, in contrast, is trained through RL to devise responses in a conversation on the basis of a particular *strategy* and in pursuit of certain objectives. In other words, ChatGPT outputs the next most likely token based on a specific context window. Cicero, in contrast, generates messages based on a plan specified by a "strategic reasoning module" (Bakhtin et al., 2022: 2). Since the demonstration of Cicero, new methods for steering (Dong et al., 2023) and controlling LLM outputs (Bhargava et al., 2023) are already being explored. For example, Bhargava et al. show that for a given LLM, there exist short prompts ("*magic words*"), which enable precise control of token generation, and thus pave the way for "the design of *LLM controllers* (programmatic or otherwise) that construct prompts on the fly to modulate LLM behaviour" (Bhargava et al., 2023: 8).

The existence of strategic reasoning modules could compel a revisiting of our understanding and usage of basic terms such as rhetoric, persuasion and propaganda to ensure they are appropriate to account for the scale, durability, and plausible intimacy of our emerging computational context. Terms such as "adversarial persuasion" and "industrialised persuasion" (Williams, 2018) come closer to conveying the level of influence over online behaviours, preferences and thoughts achievable with such systems. However, as has been argued in the pre-LLM era, the combination of surveillance, targeting, testing and automated decision making, may be more appropriately described as a form of weaponization (Nadler et al., 2018: 1) and their application along psychological and affective dimensions has been described as the "affective weaponization of information" (Davis, 2020: 2). The crucial point of differentiation between pre-digital and digital forms of persuasive content is that even the best rhetorician or state-propagandist would not be assumed to have detailed real-time information about the private lives of each individual member of their audience, nor would they be able to keep track of high-dimensional parameterisations of physiological, psychological, and environmental metrics in real-time for each individual. Nor, we add, would they be able to captivate diffuse audiences and modulate their emotional reactions in real-time with highly stimulating multimedia.

Such possibilities for manipulation are not lost on those who have developed these tools but, we argue, are downplayed. The team behind Cicero reference its manipulative potential in the article that accompanied its public announcement. "The potential danger for conversational AI agents to manipulate" they write, includes that "An agent may learn to nudge its conversational partner to achieve a particular objective" (Bakhtin et al., 2022: A.3, 3). Further, they mention that such objectives could be for or against users, for example, by teaching a new skill conversationally or defrauding someone of money. A further problem they highlight is that intents could be irrational, misinformed or involve an incitement to do harm or act unethically.

**(iii) The elicitation of how best to steer a user toward a particular objective**

The section prior to our analysis of Cicero examined several ways in which LLMs can be used to add nuance to persuasion. In this sub-section, we expand on this topic but via reference to models like Cicero. LLMs such as GPT-3 have already been shown to exhibit intrinsic capacities for persuasive communication just through prompting, without additional functionality to control outputs recursively. Jakesch et al. refer to this phenomenon as *latent persuasion* (2023). In this case, the authors investigate whether the opinions expressed by LLMs, when used as writing assistants to provide predictive text completion, influence participants' writing and attitudes. Their experimental system interpreted pauses of 750 ms as text completion requests by the user to the model, resulting in an overall 1.5s response time by the writing assistant. Importantly, their system measures parameters such as how long users pause to consider suggestions, and how many suggestions they accept.

Here, we argue that this system could be reconfigured from a latent mode of persuasion to an "active" mode by incorporating ML and RL techniques to anticipate and continuously refine suggestions until the user begins to shift their opinions. This approach would be blunt at first, but soon sharpened. As Jakesch et al. highlight, studies have shown participants may begin to change their attitudes by being encouraged to communicate beliefs contrary to their own (2023: 12). Most importantly, Jakesch et al.'s study anticipates LLMs integration into controllable dialogue models such as Cicero.

In Christopher Wylie's account of his involvement in the CA scandal, he mentions a technique in military psychological operations he refers to as "scaled perspecticide", which involves stealing and mutating a target's concept of self, and replacing it with another, through a process of deconstruction and manipulation of perception that "smothers" the target's narratives leading to domination over their information environment (Wylie, 2019). As previously mentioned, LLMs may be integrated on both sides of a conversational interface—as both the conversational agent and as part of a predictive text completion service. Such a system would represent the total enclosure of an individual in the type of information environment described by Wylie, in which their sense of self is hijacked whilst they are continuously fed information that gradually alters their perceptions and intentions.

As we have returned to throughout this article, one must consider LLM developments alongside the large-scale advertising infrastructure in which they are surely to exist. As interest in LLMs grew in 2023, platform providers revised permissions to restrict unfettered access to valuable user data via their APIs. This development has empowered existing platforms in two ways. First, by extending their domination of existing infrastructural regimes. Second, by positioning themselves as the beneficiaries of future sources of data, including data acquired through user interactions with LLMs. In each case, data gathered from user activity is preconditioned and structured within a variety of established informational enclosures. Entrapment in these enclosures makes predictive text completion a key site for directing and modulating thought by undermining user agency.

## Discussion: LLMs in New Regimes of Modulatory Power

In this conclusive section we discuss how the set of LLM capabilities outlined above could be used in concert to realise a new form of modulatory power, one that supersedes earlier forms of biopolitical disciplinary power in both intensity and scope. To illustrate this, we draw on pre-LLM work by the media-theorist David Savat, who discusses the amplification of Foucault's notion of disciplinary power through digital technologies and the emergence of a modulatory mode of power, as discussed by Deleuze. In Savat's account, disciplinary power combines hierarchical observation, normalising judgement, and examination (which combines the former two). For example, disciplinary observation is hierarchical with a high degree of immediacy between the subject and observer in a pyramidal topology of power in which the subject is aware that, at any moment, they may be being observed.

**Modulatory power and LLMs**

Modulatory power amplifies and mutates disciplinary power. In the preceding example, modulatory power would flatten observation so that it no longer has a coercive function that requires strict compliance on the part of the subject (Poster and Savat, 2009: 49). Databases, for instance, automate the observation and recording of a subject's behaviour outside of their awareness. Savat identifies four types of mechanism that underlie modulatory power, all of which may be realised through the architecture of LLM systems: the recognition of patterns; the anticipation of activity; the organisation of antitheses, which relates to the organisation of time; and the programming of flows, which unites the others. Together, these mechanisms enable modulatory control of subjects by anticipation of their actions, as individuals or groups, and by ordering or programming subjects by pre-empting moments of intervention (such as moments of indecision or vulnerability) through the capabilities of modelling and simulation, rather than coercing or moulding them into specific forms, as in the case of disciplinary power. In other words, in the case of modulatory power, the effect of control becomes so subtle that it masquerades as choice (Poster and Savat, 2009: 57).

Computational modelling and simulation are the key instruments in the modulatory mode of power that uses a "mode of observation that sees before the event" (Bogard, 1996; Poster and Savat, 2009: 47–48). Another instrument of modulatory power is categorical sorting, which involves an "infinite comparative process that determines which norms, profiles or categories you are," and underlies the generation of the patterns against which control mechanisms are continuously adjusting (Poster and Savat, 2009: 53). The final instrument highlighted by Savat is "the test or the sample," which replaces the role of examination in "determining the pattern of behaviour one exemplifies." In the case of marketing such samples provide the basis for determining patterns of consumption, leading scholars like Deleuze to consider marketing as a key form of social control (Poster and Savat, 2009: 53–54). Indeed, as noted, many aspects of modulatory power already derive from the form and function of globally connected digital infrastructure. The versality of LLM capabilities thus, when used in combination with emerging methods discussed in this article, amplify and accelerate the realisation of modulatory power.

**Modulatory power and the built environment**

The threats implied by modulatory power take on new intensity as toolsets such as LLMs transform the built environments we inhabit. As we have argued in the above, the infrastructure upon which these techniques are operationalised troubles their categorisation as tools for rhetoric or propaganda alone. Their materiality and unique affordances alter the pace and scale at which they can be used to personalise, remember, and intervene on individual and collective outcomes. Dan McQuillan summarises our broader quandary as follows:

> Data science does not affect by argument alone but acts directly in the world as a form of algorithmic force. It is machinic, that is, an assembly of flows and logic that enrolls humans and technology in a larger, purposeful structure. While algorithms and data are the bone and sinew of data science, its vital force comes from general computation. As computation becomes pervasive, capturing and reorganising human activity, data science exerts its philosophy directly as orderings, decisions and outcomes (McQuillan, 2018: 254).

Following Luitse and Denkena, we understand LLMs as "vehicles of power" in the political economy of AI whereby big tech companies have leveraged their resources to "shift power relations in their favour". Command over vast infrastructures of compute, data and derived bodies of expertise "solidifies these companies' role as rentiers in the political economy of AI" (Luitse and Denkena, 2021), and extend data colonialism in ways that echo acts of historical colonialism (Penn, 2023). As we have attended to above, LLMs also intersect with other forms of power, such as psycho-political and socio-political power, that arise from wielding their capabilities against the people, entities, objects and processes that make up the lifeworld, and which are gradually being assimilated into the infrastructural enclosures of LLM-based systems.

Precursors to these trends have been anticipated in the work of historians of technology who have documented the many ways in which the development of mathematical and statistical techniques have overlapped with the design of enclosures. "The ties between network science, urban planning, and social engineering are deeply historical, conceptual, and bi-directional," write Kurgan et. al., "Network science is haunted by the consequences of urban planning, and vice versa—smart cities are just the latest manifestation of this intricate web of influence" (Kurgan et al., 2019). Ideals about the nature and use of our physical environments have long played out through rhetoric about computational technologies, with "smartness" (Green, 2019; Halpern and Mitchell, 2022), "scalability" (Tsing, 2012) and statistical "neighborhoods" (Chun and Barnett, 2021) among the more recent examples. By this view, LLMs are simply a euphemism for a new chapter in the bitter history of modern surveillance and social engineering that visit upon the level of thought itself.

# Conclusion

This article outlines a range of societal risks posed by the rapid rise of LLMs in the 2020s, which we believe have not been adequately accounted for in the existing literature. We provide a preliminary typology of emerging examples, which we argue illustrate the role of LLMs as key components for a new regime of individual and societal control that represents a further amplification of existing regimes of modulatory power.

LLM systems may be configured as powerful interpreters of human intent, which is the basis of many applications that have been explored in 2023. However, this capability can be swiftly complemented with mechanisms for the prediction and modulation of human intent and action. LLMs may thus become the basis for powerful toolsets capable of intervening at a level of behaviour and temporality of human cognition that can gradually and adaptively direct and steer behaviour towards "strategic" objectives and even intervene on the fragile ebb and flow of our thoughts.

# References

Argyle LP, Busby EC, Fulda N, et al. (2023) Out of One, Many: Using Language Models to Simulate Human Samples. *Political Analysis* 31(3). Cambridge University Press: 337–351.

Arnold KC, Chauncey K and Gajos KZ (2020) Predictive text encourages predictable writing. In: *Proceedings of the 25th International Conference on Intelligent User Interfaces*, Cagliari Italy, 17 March 2020, pp. 128–138. ACM. Available at: https://dl.acm.org/doi/10.1145/3377325.3377523.

Bakhtin A, Brown N, Dinan E, et al. (2022) Human-level play in the game of Diplomacy by combining language models with strategic reasoning. *Science* 378(6624). American Association for the Advancement of Science: 1067–1074.

Benkler Y, Faris R and Roberts H (eds) (2018) *Network Propaganda: Manipulation, Disinformation, and Radicalization in American Politics*. Oxford University Press. Available at: https://doi.org/10.1093/oso/9780190923624.003.0009 (accessed 20 October 2023).

Bhargava A, Witkowski C, Shah M, et al. (2023) What's the Magic Word? A Control Theory of LLM Prompting. arXiv:2310.04444. arXiv. Available at: http://arxiv.org/abs/2310.04444.

Blair AM (2010) *Too Much to Know: Managing Scholarly Information before the Modern Age*. Yale University Press.

Bogard W (1996) *The Simulation of Surveillance: Hypercontrol in Telematic Societies*. Cambridge University Press.

Bommasani R, Hudson DA, Adeli E, et al. (2022) On the Opportunities and Risks of Foundation Models. arXiv:2108.07258. arXiv. Available at: http://arxiv.org/abs/2108.07258.

Bouk D (2017) The History and Political Economy of Personal Data over the Last Two Centuries in Three Acts. *Osiris* 32(1). The University of Chicago Press: 85–106.

Brand J, Israeli A and Ngwe D (2023) Using GPT for Market Research. *SSRN Electronic Journal*. Epub ahead of print 2023. DOI: 10.2139/ssrn.4395751.

Carlson D and Montgomery JM (2017) A Pairwise Comparison Framework for Fast, Flexible, and Reliable Human Coding of Political Texts. *American Political Science Review* 111(4). Cambridge University Press: 835–843.

Chiang T (2023) ChatGPT Is a Blurry JPEG of the Web. *The New Yorker*, 9 February. Available at: https://www.newyorker.com/tech/annals-of-technology/chatgpt-is-a-blurry-jpeg-of-the-web (accessed 16 March 2023).

Chu E, Andreas J, Ansolabehere S, et al. (2023) *Language Models Trained on Media Diets Can Predict Public Opinion*. arXiv:2303.16779, 28 March. arXiv. Available at: http://arxiv.org/abs/2303.16779.

Chun WHK and Barnett A (2021) *Discriminating Data: Correlation, Neighborhoods, and the New Politics of Recognition*. Cambridge, Massachusetts: The MIT Press.

Coyle D, Moore J, Kristensson PO, et al. (2012) I did that! Measuring users' experience of agency in their own actions. In: *Proceedings of the SIGCHI Conference on Human Factors in Computing Systems*, New York, NY, USA, 5 May 2012, pp. 2025–2034. CHI '12. Association for Computing Machinery. Available at: https://doi.org/10.1145/2207676.2208350.

Daston L (2022) *Rules: A Short History of What We Live By*. Princeton : Oxford: Princeton University Press.

Davis MB Elizabeth (ed.) (2020) *Affective Politics of Digital Media: Propaganda by Other Means*. New York: Routledge.

Demszky D, Yang D, Yeager DS, et al. (2023) Using large language models in psychology. *Nature Reviews Psychology*. Nature Publishing Group: 1–14.

Diaz Ruiz C (2023) Disinformation on digital media platforms: A market-shaping approach. *New Media & Society*. SAGE Publications: 14614448231207644.

Dong Y, Wang Z, Sreedhar MN, et al. (2023) SteerLM: Attribute Conditioned SFT as an (User-Steerable) Alternative to RLHF. arXiv:2310.05344. arXiv. Available at: http://arxiv.org/abs/2310.05344.

Floridi L and Chiriatti M (2020) GPT-3: Its Nature, Scope, Limits, and Consequences. *Minds and Machines* 30(4): 681–694.

Gibney E (2018) The scant science behind Cambridge Analytica's controversial marketing techniques. *Nature*. Epub ahead of print 29 March 2018. DOI: 10.1038/d41586-018-03880-4.

Goldstein JA, Sastry G, Musser M, et al. (2023) Generative Language Models and Automated Influence Operations: Emerging Threats and Potential Mitigations. arXiv:2301.04246. arXiv. Available at: http://arxiv.org/abs/2301.04246.

Green B (2019) *The Smart Enough City: Putting Technology in Its Place to Reclaim Our Urban Future*. The MIT Press.

Haggerty KD and Ericson RV (2000) The surveillant assemblage. *The British Journal of Sociology* 51(4): 605–622.

Halpern O and Mitchell R (2022) *The Smartness Mandate*. Cambridge, Massachusetts: The MIT Press. Available at: http://mitpress.mit.edu/9780262544511.

Hancock JT, Naaman M and Levy K (2020) AI-Mediated Communication: Definition, Research Agenda, and Ethical Considerations. *Journal of Computer-Mediated Communication* 25(1): 89–100.

He J, Wallis F and Rathje S (2023) *Homophily in An Artificial Social Network of Agents Powered By Large Language Models*. preprint, 24 June. In Review. Available at: https://www.researchsquare.com/article/rs-3096289/v1.

Horton JJ (2023) *Large Language Models as Simulated Economic Agents: What Can We Learn from Homo Silicus?* arXiv:2301.07543, 18 January. arXiv. Available at: http://arxiv.org/abs/2301.07543.

ICML Program Chairs (2023) LLM Policy. Available at: https://icml.cc/Conferences/2023/llm-policy (accessed 19 February 2023).

Jakesch M, Bhat A, Buschek D, et al. (2023) Co-Writing with Opinionated Language Models Affects Users' Views. In: *Proceedings of the 2023 CHI Conference on Human Factors in Computing Systems*, New York, NY, USA, 19 April 2023, pp. 1–15. CHI '23. Association for Computing Machinery. Available at: https://dl.acm.org/doi/10.1145/3544548.3581196.

Jiang H, Zhang X, Cao X, et al. (2023) PersonaLLM: Investigating the Ability of GPT-3.5 to Express Personality Traits and Gender Differences. arXiv:2305.02547. arXiv. Available at: http://arxiv.org/abs/2305.02547 (accessed 17 July 2023).

Jorge Luis Borges (n.d.) *The Library of Babel*. Available at: http://archive.org/details/TheLibraryOfBabel (accessed 27 January 2024).

Knight W (2023) Chatbot Hallucinations Are Poisoning Web Search. *Wired*, 5 October. Available at: https://www.wired.com/story/fast-forward-chatbot-hallucinations-are-poisoning-web-search/ (accessed 7 October 2023).

Kosinski M, Stillwell D and Graepel T (2013) Private traits and attributes are predictable from digital records of human behavior. *Proceedings of the National Academy of Sciences* 110(15): 5802–5805.

Kruppa BJ and M (2023) WSJ News Exclusive | ChatGPT Creator Is Talking to Investors About Selling Shares at $29 Billion Valuation. Available at: https://www.wsj.com/articles/chatgpt-creator-openai-is-in-talks-for-tender-offer-that-would-value-it-at-29-billion-11672949279 (accessed 16 March 2023).

Kurgan L, Brawley D, House B, et al. (2019) Homophily: The Urban History of an Algorithm. *e-flux*, October. Available at: https://www.e-flux.com/architecture/are-friends-electric/289193/homophily-the-urban-history-of-an-algorithm/ (accessed 31 October 2023).

Lowrey A (2023) How ChatGPT Will Destabilize White-Collar Work. *The Atlantic*, 20 January. Available at: https://www.theatlantic.com/ideas/archive/2023/01/chatgpt-ai-economy-automation-jobs/672767/ (accessed 16 March 2023).

Luitse D and Denkena W (2021) The great Transformer: Examining the role of large language models in the political economy of AI. *Big Data & Society* 8(2). SAGE Publications Ltd: 20539517211047734.

Marche S (2022) The College Essay Is Dead. Available at: https://www.theatlantic.com/technology/archive/2022/12/chatgpt-ai-writing-college-student-essays/672371/ (accessed 16 March 2023).

Marlin R (2002) *Propaganda and the Ethics of Persuasion*. Ontario: Broadview Press.

Matz S, Teeny J, Vaid SS, et al. (2023) The Potential of Generative AI for Personalized Persuasion at Scale. OSF. Epub ahead of print 31 December 2023. DOI: 10.31234/osf.io/rn97c.

Matz SC, Kosinski M, Nave G, et al. (2017) Psychological targeting as an effective approach to digital mass persuasion. *Proceedings of the National Academy of Sciences* 114(48): 12714–12719.

McQuillan D (2018) Data Science as Machinic Neoplatonism. *Philosophy & Technology* 31(2): 253–272.

Meta (2023) Introducing New AI Experiences Across Our Family of Apps and Devices. In: *Meta*. Available at: https://about.fb.com/news/2023/09/introducing-ai-powered-assistants-characters-and-creative-tools/ (accessed 20 December 2023).

Meta AI (2022) CICERO: An AI agent that negotiates, persuades, and cooperates with people. Available at: https://ai.facebook.com/blog/cicero-ai-negotiates-persuades-and-cooperates-with-people/ (accessed 16 March 2023).

Mills S and Sætra HS (2022) The autonomous choice architect. *AI & SOCIETY*. Epub ahead of print 22 June 2022. DOI: 10.1007/s00146-022-01486-z.

Morgan MS (2012) *The World in the Model: How Economists Work and Think*. Cambridge University Press.

Nadler A, Crain M and Donovan J (2018) *The Political Perils of Online Ad Tech*. 17 October. Data & Society. Available at: https://datasociety.net/library/weaponizing-the-digital-influence-machine/.

Naughton J (2022) I wrote this column myself, but how long before a chatbot could do it for me? *The Observer*, 10 December. Available at: https://www.theguardian.com/commentisfree/2022/dec/10/i-wrote-this-column-myself-but-how-long-before-a-chatbot-could-do-it-for-me (accessed 16 March 2023).

Nay JJ (2023) Large Language Models as Corporate Lobbyists. arXiv:2301.01181. arXiv. Available at: http://arxiv.org/abs/2301.01181.

Nixon R (2011) Slow Violence and the Environmentalism of the Poor. In: *Slow Violence and the Environmentalism of the Poor*. Harvard University Press.

*NVIDIA Keynote at SIGGRAPH 2023* (2023). Available at: https://www.youtube.com/watch?v=Z2VBKerS63A (accessed 20 December 2023).

*OpenAI DevDay, Opening Keynote* (2023). Available at: https://www.youtube.com/watch?v=U9mJuUkhUzk (accessed 14 November 2023).

Ornstein JT, Blasingame EN and Truscott JS (2023) How to Train Your Stochastic Parrot: Large Language Models for Political Texts. Epub ahead of print July 2023.

Park JS, O'Brien JC, Cai CJ, et al. (2023) *Generative Agents: Interactive Simulacra of Human Behavior*. arXiv:2304.03442, 6 April. arXiv. Available at: http://arxiv.org/abs/2304.03442.

Park PS, Schoenegger P and Zhu C (2023) Diminished Diversity-of-Thought in a Standard Large Language Model. arXiv:2302.07267. arXiv. Available at: http://arxiv.org/abs/2302.07267.

Penn J (2023) Animo nullius: on AI's origin story and a data colonial doctrine of discovery. *BJHS Themes*. Cambridge University Press: 1–16.

Pickering MJ and Garrod S (2013) An integrated theory of language production and comprehension. *The Behavioral and Brain Sciences* 36(4): 329–347.

Poerksen U (1995) *Plastic Words: The Tyranny of a Modular Language*. University Park, Pa: Penn State University Press.

Poster M and Savat D (2009) *Deleuze and New Technology*. Deleuze connections. Edinburgh: Edinburgh University Press.

Rosenbusch H, Stevenson CE and van der Maas HLJ (2023) How Accurate are GPT-3's Hypotheses About Social Science Phenomena? *Digital Society* 2(2): 26.

Serapio-García G, Safdari M, Crepy C, et al. (2023) *Personality Traits in Large Language Models*. preprint, 28 August. In Review. Available at: https://www.researchsquare.com/article/rs-3296728/v1.

Seshia SA, Sadigh D and Sastry SS (2020) Towards Verified Artificial Intelligence. arXiv:1606.08514. arXiv. Available at: http://arxiv.org/abs/1606.08514.

Sharp B, Danenberg N and Bellman S (2018) Psychological targeting. *Proceedings of the National Academy of Sciences* 115(34).

Sommart T (2023) Formal Verification of Neural Networks. Available at: https://cds.cern.ch/record/2867415.

Staab R, Vero M, Balunović M, et al. (2023) Beyond Memorization: Violating Privacy Via Inference with Large Language Models. arXiv:2310.07298. arXiv. Available at: http://arxiv.org/abs/2310.07298.

Stark L (2018) Algorithmic psychometrics and the scalable subject. *Social Studies of Science* 48(2): 204–231.

Synthetic Users (n.d.). Available at: https://www.syntheticusers.com/ (accessed 17 March 2024).

Tapus A, Țăpuș C and Matarić MJ (2008) User—robot personality matching and assistive robot behavior adaptation for post-stroke rehabilitation therapy. *Intelligent Service Robotics* 1(2): 169–183.

Tiku N (2017) Welcome to the Next Phase of the Facebook Backlash. *Wired*, 21 May. Available at: https://www.wired.com/2017/05/welcome-next-phase-facebook-backlash/ (accessed 10 February 2024).

Tsing AL (2012) On Nonscalability. *Common Knowledge* 18(3): 505–524.

Veselovsky V, Ribeiro MH and West R (2023) Artificial Artificial Artificial Intelligence: Crowd Workers Widely Use Large Language Models for Text Production Tasks. arXiv:2306.07899. arXiv. Available at: http://arxiv.org/abs/2306.07899.

Wartofsky MW (1979) Models, Metaphysics and the Vagaries of Empiricism. In: Wartofsky MW (ed.) *Models: Representation and the Scientific Understanding*. Boston Studies in the Philosophy of Science. Dordrecht: Springer Netherlands, pp. 24–39.

Weidinger L, Uesato J, Rauh M, et al. (2022) Taxonomy of Risks posed by Language Models. In: *2022 ACM Conference on Fairness, Accountability, and Transparency*, New York, NY, USA, 20 June 2022, pp. 214–229. FAccT '22. Association for Computing Machinery. Available at: https://doi.org/10.1145/3531146.3533088.

Whittaker M (2021) The steep cost of capture. *Interactions* 28(6): 50–55.

Wylie C (2019) *Mindf*ck: Cambridge Analytica and the Plot to Break America*. Random House.

Yeung K (2017) 'Hypernudge': Big Data as a mode of regulation by design. *Information, Communication & Society* 20(1): 118–136.

Zittrain J (2022) Intellectual Debt: With Great Power Comes Great Ignorance. In: Mueller O, Kellmeyer P, Voeneky S, et al. (eds) *The Cambridge Handbook of Responsible Artificial Intelligence: Interdisciplinary Perspectives*. Cambridge Law Handbooks. Cambridge: Cambridge University Press, pp. 176–184.

Zollmann F (2017) Bringing Propaganda Back into News Media Studies. *Critical Sociology*: 089692051773113.